\documentclass[aps,prl,twocolumn,superscriptaddress]{revtex4}
\usepackage{graphicx}

\newcommand{\eps} {\varepsilon}

\begin{document}

\title{Phase-separation transition in liquid mixtures near curved
charged objects}

\author{Gilad Marcus}
% \email{gilad.marcus@mpq.mpg.de}
\affiliation{Max-Planck-Institut f\"{u}r Quantenoptik, Hans-Kopfermann-Str. 1, D-85748
Garching, Germany.}

\author{Sela Samin}
% \email{samins@bgu.ac.il}
\affiliation{Department of Chemical Engineering, Ben-Gurion University of the Negev,
84105 Beer-Sheva, Israel.}

\author{Yoav Tsori}
% \email{tsori@bgu.ac.il}
\affiliation{Department of Chemical Engineering, Ben-Gurion University of the Negev,
84105 Beer-Sheva, Israel.}

\date{30/7/2008}

\begin{abstract}

We study the thermodynamic behavior of nonpolar liquid mixtures in the vicinity of
curved charged objects, such as electrodes or charged colloids.
There is a critical value of charge (or potential), above which a
phase-separation transition occurs, and the interface between high- and low-dielectric
constant components becomes sharp.
Analytical and numerical composition profile are given, and the equilibrium front location
as a function of charge or voltage is found.
We further employ a simple Cahn-Hilliard type equation to study the dynamics of phase-separation
in spatially nonuniform electric fields. We find an exponential temporal relaxation of
the demixing front location. We give the dependence of the steady-state location and
characteristic time on the charge, mixture
composition and ambient temperature.

\end{abstract}

% \pacs{61.25.Hq, 64.70.Nd, 61.41.+e}

\maketitle

Situations where charged objects, such as electrodes or colloids, are found in liquid
environments are abundant in science and technology. In ionic mixtures, experiments and
theory show that ions migrate toward the object and lead to screening of the electric
field.
In nonpolar liquids, the situation is different: the decay of electric field far
from the objects depends on the geometry of all conducting surfaces, and may be long
range. When a nonpolar liquid mixture is under the influence of a uniform electric field
$E$, the
theories of Landau \cite{LL1} and later Onuki \cite{onuki1} showed that the
critical temperature can change by a small amount, proportional to $E^2$. Experiment by
Debye and Kleboth \cite{debye} partially confirmed the theory.

However, here we show that the situation in spatially nonuniform electric fields, occurring
when liquid mixtures are found under the influence of curved charged surfaces,
is quite different.
When the temperature $T$ is larger than the critical temperature
$T_c$,
the mixture exhibits smooth composition variations. This
dielectrophoretic
behavior is reminiscence of the effect of gravity.\cite{moldover_rmp79}
For a homogeneous mixture below $T_c$,
there are two scenarios: if the charge density is small,
there are still weak composition gradients. On the other hand,
large enough charge leads to a
phase-separation transition, where the liquid with high-dielectric constant is
close
to the high field region while the liquid with low dielectric constant is pulled
away, and the coexisting domains are separated by a sharp composition front.

The phase transition described below occurs in systems described by
bistable free-energy functionals giving rise to a phase diagram in
the composition-temperature plane divided into two regions:
homogeneous mixture and a phase-separated state. In order to be
specific and to facilitate the connection to experiment, we consider
the following binary mixture free-energy density
$f_m=kT\tilde{f}_m/Nv_0$, where
\begin{eqnarray}\label{fm}
\tilde{f}_m=\phi\log(\phi)+
(1-\phi)\log(1-\phi)+N\chi\phi(1-\phi)
\end{eqnarray}
This symmetric ($N_A=N_B=N$) free energy is given in terms of the A-component
composition
$\phi$ ($0<\phi<1$) in a mixture of A/B liquids, and the so-called Flory parameter
$\chi\sim 1/T$.\cite{doi} Simple liquids have
$N=1$, while polymers have $N>1$ monomers, each of volume $v_0$.
$k$ is the Boltzmann constant.
The critical point is given at $(\phi_c,(N\chi)_c)=(1/2,2)$. In the absence of electric field,
the mixture is homogeneous if $T>T_t$, and
unstable otherwise. The transition (binodal) temperature $T_t$
at a given composition is given by
$T_t=(N\chi)_cT_c\left[\log(\phi/(1-\phi))/(2\phi-1)\right]^{-1}$.\cite{doi}
The phase transition does not depend on the exact form of $f_m$, and
appears in a Landau series expansion of Eq. (\ref{fm}) around $\phi_c$, or in any
other similar ``double-well'' free-energy functional.

As is shown below, the effect of electric fields is large if they
originate from curved charged surfaces. In this work we consider
for simplicity surfaces with fixed curvature: a charged spherical
colloid, a charged wire or two concentric cylinders, and the
``wedge'' capacitor, made up from two flat and nonparallel surfaces.
Fixed charges on the conductors, fixed potentials, or a combination
of the two are considered by us. When the mixture is in the
vicinity of a charged object with a fixed surface charge, the total
dimensionless free-energy is $\tilde{f}=\tilde{f}_m+\tilde{f}_{\rm es}$, where
$\tilde{f}_{\rm es}=(Nv_{0}/kT)[(1/2)\eps(\phi)(\nabla\psi)^2]$ is
the dimensionless electrostatic free energy density.\cite{LL1,onuki2} Note that
we do not include any direct short- or long-range interactions between
the liquid and the confining walls.

The equilibrium state is a solution of the two coupled nonlinear
equations: $\delta \tilde{f}/\delta\phi=0$ and
$\delta \tilde{f}/\delta\psi=0$, where $\psi$ is the
electrostatic potential obeying the proper boundary
conditions.\cite{onuki1,TTL} The equation $\delta
\tilde{f}/\delta\psi=0$ leads to Laplace's equation:
$\nabla\cdot(\eps(\phi)\nabla\psi)=0$, and is readily solved by the use of Gauss' law for
systems with
prescribed charges on the confining conductors and in azimuthal or
spherical symmetries. For example, for a mixture confined
between two infinite concentric cylinders of radii $R_1$ and $R_2>R_1$, we
find $E=-\nabla\psi=\lambda/(2\pi\eps(\phi)r)$, where $\lambda$ is the charge per unit
length on the inner
cylinder and $r$ is the distance from the cylinder's center. Subsequently,
$\delta\tilde{f}_{\rm es}/\delta\phi=-N^2v_0\chi/(kT_c)(\lambda/(4\pi\eps
r))^2d\eps/d\phi$.
Similarly, $E=Q/(4\pi\eps(\phi) r^2)$ for a spherical colloid of radius $R_1$ and
charge $Q$, and $r$ is the distance from the colloid's center, and $E=V/(\beta r)$ for a
wedge consisting of two flat conductors with potential difference $V$ and opening angle
$\beta$ between them, and $r$ is the distance from the imaginary meeting point of the
conductors.

We thus arrive at a considerable simplification of the problem, since
the expression for $E$ obtained above allows to write a single dimensionless governing
equation for all three cases with radial or azimuthal symmetry:
\begin{eqnarray}\label{gov_eqn}
\log\left(\frac{\phi}{1-\phi}\right)+N\chi(1-2\phi) -N\chi
M\frac{d\tilde{\eps}/d\phi}{\tilde{\eps}^2(\phi)}\tilde{r}^{-n}
-\mu&=&0~~~~~~
\end{eqnarray}
In the above, $M$
is the dimensionless ratio between the maximum electrostatic energy stored in a
molecular
volume and the thermal energy.
$M$ is $M_c\equiv  \lambda^2Nv_0/(16\pi^2kT_cR_1^2\eps_0)$ for two concentric
 cylinders,
$M$ is $M_s\equiv Q^2Nv_0/(64\pi^2kT_c\eps_0R_1^4)$
for a spherical colloid, and
$M$ is $\tilde{\eps}^2M_w$ for the wedge, where
$M_w\equiv V^2Nv_0\eps_0/(4\beta^2kT_cR_1^2)$, $V$ is the voltage between the
wedge plates, and $R_1$ is the smallest distance from the
conductors' edge to their imaginary meeting point.
$\tilde{r}\equiv r/R_1$ is the scaled distance
from the center of the sphere or the inner cylinder, and $\tilde{\eps}=\eps/\eps_0$,
where $\eps_0$ is the vacuum permittivity.
Finally, $n$ is the exponent characterizing the fall of $E^2$: $n=2$ for
concentric cylinders and the wedge geometries, and $n=4$ for the sphere.
The importance of curvature is
exemplified by the appearance of $R_1$ in the expressions for the $M$'s.
$\mu$ is a Lagrange multiplier needed to conserve the
average mixture composition: $\langle \phi({\bf r})\rangle=\phi_0$, and $\phi_0$
is the average composition. In the case of an open system coupled to a particle
reservoir at $r\to\infty$, $\mu$ is the reservoir's chemical potential.
The phase-transition described below is from a homogeneous (mixed) to a demixed state,
and
therefore it is assumed that $\phi_0$ is outsides of the binodal curve,
namely, $T>T_t$.
\begin{figure}[!th]
\begin{center}
\includegraphics[scale=0.45,bb=64 205 520 560,clip]{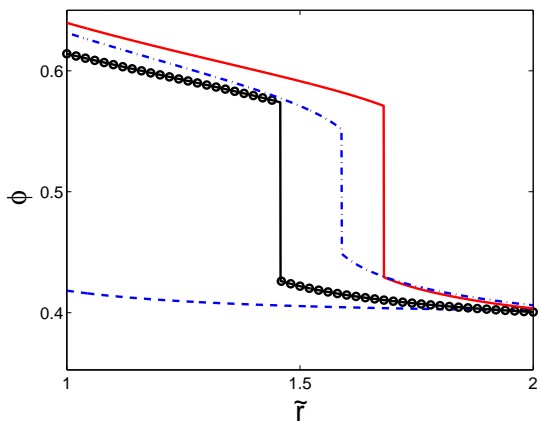}
\end{center}
\caption{
Equilibrium profiles $\phi(\tilde{r})$ for concentric cylinders with different
dimensionless charge $M_c$.
Dashed line: $T=0.991T_c$, and $M_c=0.008$
too small for phase separation. Circles: same $T$, but $M_c=0.04$.
Solid line: same $T$, but
$M_c=0.08$. Dash-dot line: $M_c=0.08$, but $T=0.994T_c$ is a higher temperature.
We took
fixed average composition $\phi_0=0.4$. In this and other figures,
$\tilde{R}_1=1$, $\tilde{R}_2=5$, $\eps_a=5\eps_0$, and $\eps_b=3\eps_0$.
}
\end{figure}

Equation (\ref{gov_eqn}) expresses implicitly the composition profile
$\phi(\tilde{r})$.
Above $T_c$, $\phi(\tilde{r})$ has only smooth
variations, irrespective of the value of $M$.
Below $T_c$ [equivalently $N\chi>(N\chi)_c$],
if $M$ is sufficiently small,
the profile $\phi(\tilde{r})$ is smooth, with high-$\phi$ values at small $\tilde{r}$'s
and low values at
larger radii.
\begin{figure}[th!]
\begin{center}
\includegraphics[scale=0.55,bb=125  110 485 660,clip]{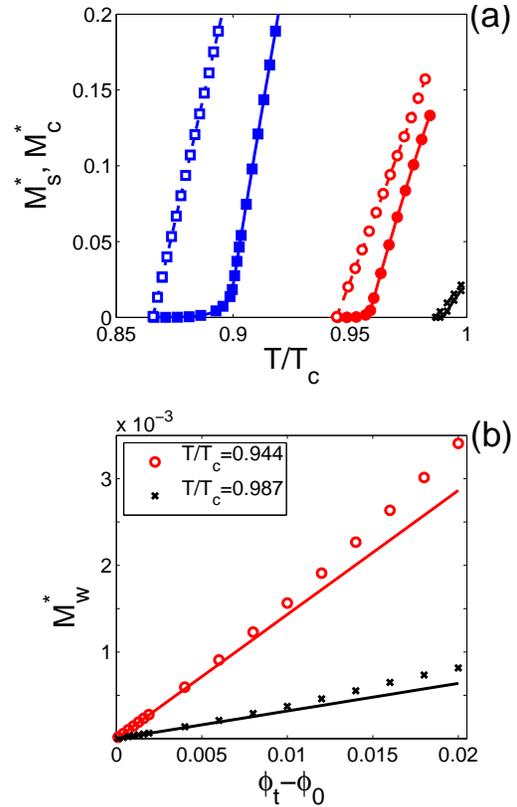}
\end{center}
\caption{(a) Filled symbols: dimensionless critical charge $M_s^*$ for
phase-separation
near an isolated spherical colloid as a function of temperature. The colloid is
coupled
to a reservoir with three compositions: $\phi_0=0.2$ (squares), $0.3$
(circles) and $0.4$ (crosses). Open symbols: $M_c^*$ for a closed cylindrical
system with same compositions.
(b) $M_w^*$ vs $\Delta\phi=\phi_t-\phi_0$ from
Eq. (\ref{Mw_star_analytic}) (solid line) and from numerics (symbols).
}
\end{figure}
However, there is a critical
value of $M$, denoted $M^*$, above which $\phi(\tilde{r})$ exhibits a sharp
jump: for
$M>M^*$,
high- and low-$\phi$ domains coexist separated by a clear interface at $\tilde{r}=\tilde{R}$.
This transition occurs generally,
even when the constitutive relation $\eps(\phi)$ is linear in $\phi$.
This is in contrast to the
Landau mechanism, which relies on a quadratic dependence of $\eps$ on $\phi$ and
is responsible to a small change in $T_c$.
We therefore chose
the linear relation $\eps(\phi)=\eps_b+\phi\Delta\eps$, where
$\Delta\eps\equiv\eps_a-\eps_b$, and $\eps_a$ and $\eps_b$ are the dielectric
constants of components A and B, respectively.

The typical demixing electric fields and surface or line charge density can be
estimated
from the values of $M$ (see Figs. 1 and 2). At $M_s=0.001$ and using a molecular volume
of
$Nv_0=10^{-26}$ m$^3$, colloid's radius $R_1=1~\mu$m, $T_c\simeq 300$ K,
 $\tilde{\eps}\simeq 4$, we find the electric field at the sphere's edge to be
$E\sim 10^6-10^7$ V/m (surface potential $\sim 1-10$ V).
The corresponding charge density is $\sigma=\eps E\sim 10^{-5}-10^{-4}$ C/m$^2$
(total charge $Q=800-8000e$). Similar values for the electric
field and charge density appear in the concentric cylinders and wedge geometries.

Figure 1 shows $\phi(\tilde{r})$ for a binary mixture confined by two concentric
cylinders for several values of the dimensionless parameter $M_c$ and at two
different temperatures. When
$M_c=0.008$, there is no phase separation, and the profile is
smooth. As $M_c$ increases above $M_c^*$,
phase separation occurs, and $\phi(\tilde{r})$ rapidly changes from high to low
values at
the phase-separation front located at $\tilde{r}=\tilde{R}$.
Further increase of $M_c$ at constant temperature leads to displacement of $\tilde{R}$
to larger values and to larger composition difference between
coexisting domains.\cite{KYL,TL_CRphysique}

Figure 2(a) shows the calculated critical value $M_s^*$ as a
function of temperature for a spherical colloid coupled to a
particle reservoir at three different compositions. At a given $T$
above $T_t$, larger values of $|\phi_c-\phi_0|$ require more charge
for demixing. Curves also show  $M_c^*$ for a system enclosed between
two concentric cylinders. Notice that approaching $T_t$, $M^*$ becomes
infinitesimally small.
For a wedge with average composition $\phi_0$ close to the
transition composition $\phi_t$ at given temperature, we obtain the
following approximation:
\begin{eqnarray}\label{Mw_star_analytic}
M_w^*=\frac{\phi_t-\phi_0}{4\Delta\tilde{\eps}}\frac{T}{T_c}\frac{d^2\tilde{f}
_m(\phi_t)}{d\phi^2}g(x)
\end{eqnarray}
where $x\equiv R_2/R_1$ and $g(x)=2(x^2-1)/(x^2-1-2\ln x)$.
Figure 2(b) shows $M_w^*$ from this formula and compares it with a more accurate
numerical solution.
\begin{figure}[th!]
\begin{center}
\includegraphics[scale=0.4,bb=15 240 550 600,clip]{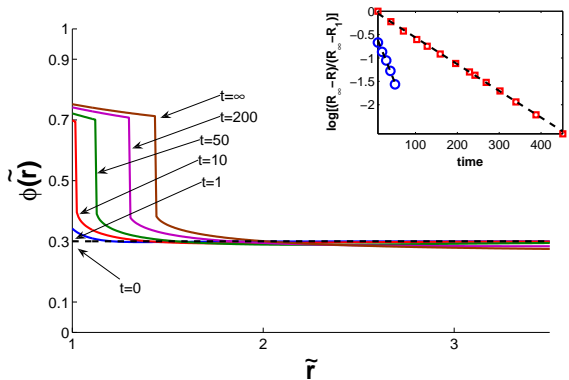}
\end{center}
\caption{Composition profiles $\phi(\tilde{r},t)$ at several dimensionless times
for concentric cylinders with $M_c=0.32$, $\phi_0=0.3$, and $T=0.95T_c$.
Inset: semilog plot of $\tilde{R}(t)$. Numerical results (squares) and
experiments of Ref. \cite{TTL} (circles, time in s).
}
\end{figure}

We now turn to describe the relaxation toward equilibrium.
The dynamics are governed by the following set of equations:\cite{onuki3,tanaka,bray}
\begin{eqnarray}
\frac{\partial\phi}{\partial t}+{\bf u}\cdot\nabla\phi&=&L\nabla^2\delta
f/\delta\phi\label{1st} \\
\nabla\cdot(\eps(\phi)\nabla\psi)&=&0\label{2nd}\\
\nabla \cdot{\bf u}&=&0\label{3rd}\\
\rho\left[\frac{\partial {\bf u}}{\partial t}+({\bf u}\cdot{\bf \nabla}){\bf
u}\right]&=&\eta\nabla^2{\bf u}-\nabla P-\phi\nabla\delta f/\delta\phi\label{4th}
\end{eqnarray}
${\bf u}$ is the velocity field corresponding
to hydrodynamic flow and $\eta$ is the liquid viscosity. Equation (\ref{1st}) is a
continuity
equation for $\phi$, where $-L\nabla(\delta f/\delta \phi)$ is the diffusive
current due to the inhomogeneities of the chemical potential, and $L$ is the
transport coefficient (assumed constant). Equation (\ref{2nd}) is Laplace's
equation, Eq. (\ref{3rd}) implies incompressible flow, and Eq. (\ref{4th}) is
Navier-Stokes equation with a force term $-\phi\nabla\delta
f/\delta\phi$.\cite{bray}

We continue in the limit of overdamping and with the assumption of azimuthal
symmetry. It follows that ${\bf u}= 0$. We use the dimensionless time
$\tilde{t}=Nv_0R_1^2 t/(LkT)$,
radius $\tilde{r}=r/R_1$ and energy $\tilde{f}=Nv_0f/kT$, to express $\phi$ as
a solution to a diffusionlike equation $\partial\phi/\partial t=\nabla^2\delta
f/\delta\phi$, while satisfying Laplace's equation, where the ``$\sim$'' signs have
been omitted for brevity of notation.
The time dependence of
the profile $\phi(\tilde{r},t)$, obtained numerically, is shown in Fig. 3 for several
times $t$.

\begin{figure}[th!]
\begin{center}
\includegraphics[scale=0.45,bb=28 220 530 600,clip]{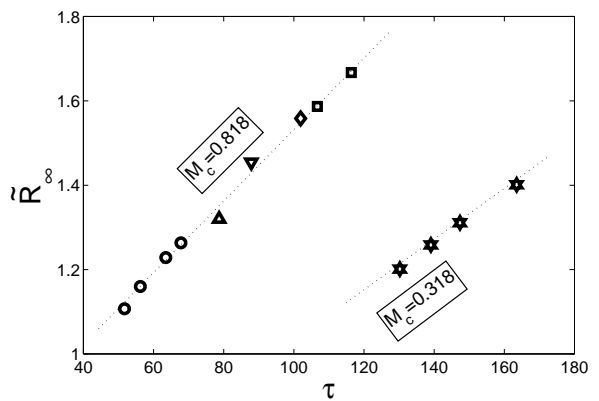}
\end{center}
\caption{
Steady-state front location $\tilde{R}_\infty$ vs time constant $\tau$ for
different
 temperatures
 and compositions. Stars: $M_c=0.318$, $\phi_0=0.3$ and $0.95\leq T/T_c\leq
0.99$. Circles: $M_c=0.818$, $\phi_0=0.2$, and $0.89\leq T/T_c\leq 0.95$.
Squares: $M_c=0.818$, $\phi_0=0.3$, and $0.95\leq T/T_c\leq 0.97$.
Up and down triangles and diamond: $\phi_0=0.22$, $\phi_0=0.24$, and
$\phi_0=0.26$,
respectively, $M_c=0.818$ and $T/T_c=0.95$.
}
\end{figure}

The dimensionless location of the demixing front
changes as a function of time: $\tilde{R}=\tilde{R}(t)$ and asymptotically tends
toward the steady-state front location $\tilde{R}_\infty$ at long times. We find
excellent match with an exponential relaxation of the form
$\tilde{R}(t)=\tilde{R}_\infty+(1-\tilde{R}_\infty)\exp(-t/\tau)$, as is shown
in the inset of Fig. 3.

Figure 4 shows the location of the steady-state demixing front
$\tilde{R}_\infty$ and the time constant $\tau$ at several different
values of $\phi_0$ and $T$, and for two different values of $M_c$.
It is worth noting that all the points with the same $M_c$ seem to fall
on the same line.  Similarly, the dependence
of $\tilde{R}_\infty$ on $\phi_0$ is displayed in Fig. 5(a).
Clearly, the domain size increases with $M_c$ at constant
temperature and composition. Increase of $\phi_0$ at constant $T$
and $M_c$ increases the domain size. Figure 5(b) shows how $\tau$
depends on $\phi_0$. Compositions closer to $\phi_c$ exhibit slower
relaxations. In addition, increase of $M_c$ leads to faster
relaxation toward steady state.

It should be emphasized that this phase transition is not restricted
to the vicinity of the critical point, and it occurs at all compositions,
provided that the electric field is large enough.
Moreover, field-induced prewetting could also be realized in vapor-liquid systems of
pure substances subject to nonuniform electric fields.
\begin{figure}[th!]
\begin{center}
\includegraphics[scale=0.5,bb=32 175 515 645,clip]{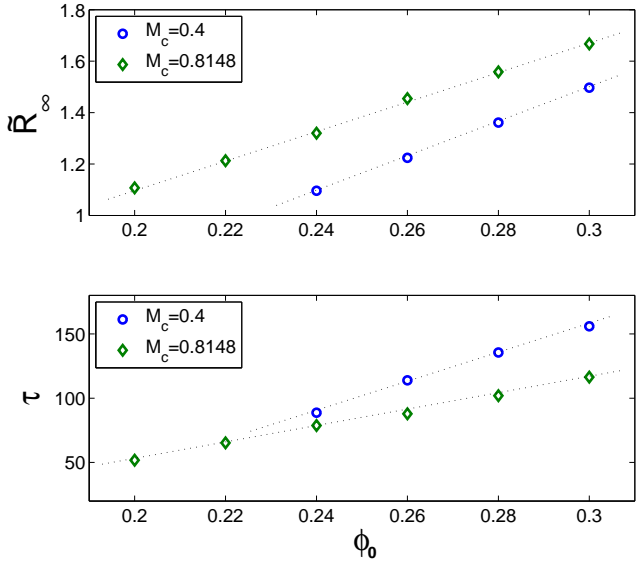}
\end{center}
\caption{Steady-state front location $\tilde{R}_\infty$ (a) and time constant
$\tau$  (b) vs $\phi_0$ for
two values of dimensionless charge $M_c$. Numerical solution for
concentric cylinders with $\tilde{R}_1=1$, $\tilde{R}_2=5$, and $T=0.95T_c$.
}
\end{figure}

There are several circumstances where the field-induced separation may have an
important influence. Colloidal suspension in liquid mixtures and polymer solutions have
been extensively studied. \cite{frenkel,lekker,larson,beysens,bechinger}
We point out that standard wetting theory is insufficient to describe these
experiments if the colloids are charged. The enrichment
layer around the colloid is sensitive to the colloid's charge, and
this may have an effect on the intercolloid interaction and hence on the
phase behavior and the rheology of suspensions.\cite{beysens,bechinger}

A drastic change to the rheological properties is also predicted for a mixture confined,
for example,
between two rotating coaxial cylinders (Taylor-Couette flow). The classic (zero field)
flow
profile would change markedly if a potential is imposed between the two cylinders.
Once the homogeneous mixture demixes,
most of the velocity gradient will fall on the liquid component with smaller
viscosity.\cite{TL_pnas} A change to the lubrication in microelectromechanical
systems
and in microfluidic channels can be similarly brought by the application of external
potential, recalling that in these systems the electric field is inherently
nonuniform.

Lastly, we point out that the demixing transition creates optical interfaces, since the
mixture's components have different refraction indices. Consequently, the propagation
of a light beam through a mixture in a channel will be altered once an
electric field creates optical interfaces, and this may be used to scatter,
focus, or even guide rays in microfluidic arrays.\cite{quake_review}

We thank L. Leibler and F. Tournilhac for help in developing
the ideas presented in this work, and D. Andelman for numerous useful comments.
This research was supported by the Israel Science
foundation (ISF) grant no. 284/05, and by the German Israeli Foundation (GIF) grant
no. 2144-1636.10/2006.

%------------------------------------------------

\end{document}